\documentclass[italian,english]{article}
\usepackage[T1]{fontenc}
\usepackage[latin9]{inputenc}
\usepackage{color}

\makeatletter
\newcommand{\lyxaddress}[1]{
\par {\raggedright #1
\vspace{1.4em}
\noindent\par}
}

\makeatother

\usepackage{babel}
\begin{document}

\title{\textbf{General relativity and OPERA Experiment}}

\author{\textbf{Christian Corda}}

\maketitle
\begin{center}
Institute for Theoretical Physics and Advanced Mathematics Einstein-Galilei,
Via Santa Gonda 14, 59100 Prato, Italy 
\par\end{center}

\lyxaddress{\begin{center}
\textit{E-mail addresses:} \textcolor{blue}{cordac.galilei@gmail.com} 
\par\end{center}}
\begin{abstract}
In his paper ``A very simple solution to the OPERA neutrino velocity
problem'' the author J. Manuel Garcia-Islas claims to have very easily
solved and explained within the general theory of relativity that
OPERA's neutrinos are not traveling faster than the speed of light
and the early time arrival is due to the presence of the Earth's gravitational
field. In this letter we easily show that the argument by Garcia-Islas
does not work.

Although it looks that data suggesting that neutrinos can travel faster
than light probably resulted from a faulty connection in a GPS timing
system, it is important to clarify that, in any case, the general
relativistic effect discussed by Garcia-Islas cannot explain the original
OPERA's data.
\end{abstract}
\textbf{PACS numbers: 04.20.-q, 14.60.Lm.}

The OPERA collaboration claimed that \textquotedbl{}the measurement
indicates an early arrival time of CNGS muon neutrinos with respect
to the one computed assuming the speed of light in vacuum\textquotedbl{}
\cite{key-1}. Such a claim generated an interesting debate within
the Scientific Community. Various authors try to justify or invalidate
the results of OPERA by using theoretical analyses, other authors
discuss potential errors in OPERA's experimental methodology, especially
concerning the clocks' synchronization, see for example refs. \cite{key-2,key-3,key-4,key-5,key-6,key-7}.
Now, it looks that data suggesting that neutrinos can travel faster
than light probably resulted from a faulty connection in a GPS timing
system \cite{key-13}.

In \cite{key-8} the author claims to have very easily solved and
explained within the general theory of relativity that OPERA's neutrinos
are not traveling faster than the speed of light and the early time
arrival is due to the presence of the Earth's gravitational field.
In this letter we easily show that the argument in \cite{key-8} does
not work. In fact, although the original OPERA's data could be wrong,
see for example \cite{key-13}, it is important to clarify that, in
any case, the general relativistic effect discussed in \cite{key-8}
cannot explain such original OPERA's data.

In \cite{key-8} the problem is simplified by assuming that on Earth
a massive particle is traveling at velocity $v$ in a circular orbit
(or just in an arc $\triangle\varphi$) at a fixed radial distance
$r=R$. Then, the proper time of travel measured by an observer which
is fixed at the same radial distance is calculated \cite{key-8}.
The analysis is performed by using the well known Schwarzschild solution
to Einstein's field equation \cite{key-8}. A first comment is needed.
Actually, general relativistic effects, i.e. the presence of the gravitational
field, are taken into due account in the process of clocks' synchronization
when the GPS receivers are used \cite{key-9}. This is exactly the
case of the OPERA experiment, which worked with two identical systems
installed at CERN and LNGS and composed of a GPS receiver for time-transfer
applications Septentrio PolaRx2e operating in \textquotedblleft{}common-view\textquotedblright{}
mode and a Cs atomic clock Symmetricom Cs4000 \cite{key-1}. More,
synchronization of GPS does not work in Schwarzschild coordinates.
Indeed, it works in the Universal Coordinated Time as maintained by
the U.S. Naval Observatory on the rotating geoid, but with synchronization
established in an underlying, locally inertial, reference frame \cite{key-9}.
In such a reference frame, by setting $G=c=1$ with the sign conventions
for the line element $(-,+,+,+),$ an approximate solution of Einstein's
field equations in isotropic coordinates is used \cite{key-9}

\begin{equation}
ds^{2}=-\left(1+2(V-\Phi_{0})\right)dt^{2}+\left(1-2V\right)\left(dr^{2}+r^{2}(\sin^{2}\theta d\varphi^{2}+d\theta^{2})\right)\label{eq: sol. approssimata}
\end{equation}

where 

\begin{equation}
V=-\frac{M}{r}\left[1-J_{2}\left(\frac{a_{1}}{r}\right)^{2}P_{2}\cos\theta\right],\label{eq: V}
\end{equation}

\begin{equation}
\Phi_{0}=-\left(\frac{M}{a_{1}}+\frac{MJ_{2}}{2a_{1}}+\frac{1}{2}\omega^{2}a_{1}^{2}\right).\label{eq: fi zero}
\end{equation}
$M$ is the Earth's mass, $J_{2}$ is Earth\textquoteright{}s quadrupole
moment coefficient, $a_{1}$ is Earth's equatorial radius and $\omega$
the Earth's angular velocity. 

In any case, we start by showing that the conclusions in \cite{key-8}
are not correct in the framework of the Schwarzschild solution. Then,
we will discuss the coordinates (\ref{eq: sol. approssimata}) too. 

Let us review the analysis in \cite{key-8}.

The Schwarzschild line element reads \cite{key-8,key-10} (but see
\cite{key-11} for clarifying historical notes to this notion)

\begin{equation}
ds^{2}=-(1-\frac{2M}{r})dt^{2}+\frac{dr^{2}}{1-\frac{2M}{r}}+r^{2}(\sin^{2}\theta d\varphi^{2}+d\theta^{2}),\label{eq: Hilbert}
\end{equation}

being $M$ the Earth's mass.

Hence, the world line for a massive particle which travels in a circular
orbit $r=R$ at a velocity $v$ in the space-time of eq. (\ref{eq: Hilbert})
is \cite{key-8}

\begin{equation}
x^{\mu}(\tau)=\left(\gamma\tau,\mbox{ }R,\mbox{ }\frac{\pi}{2},\mbox{ }\frac{\gamma v\tau}{R}\right),\label{eq: worldline}
\end{equation}

where 
\begin{equation}
\gamma=\frac{1}{\sqrt{1-v^{2}-\frac{2M}{R}}}\label{eq: gamma factor}
\end{equation}

is the general \textquotedblleft{}gamma factor\textquotedblright{}
in presence of a gravitational field and $\tau$ is the proper time
for the world line (\ref{eq: worldline}). The gamma factor (\ref{eq: gamma factor})
is obtained by imposing the 4-velocity to be orthogonal for the world
line \cite{key-8,key-10}

\begin{equation}
g_{\mu\nu}\frac{dx^{\mu}}{d\tau}\frac{dx^{\nu}}{d\tau}=-1.\label{eq: 4-velocity}
\end{equation}

Eq. (\ref{eq: worldline}) gives \cite{key-8}

\begin{equation}
\frac{d\varphi}{d\tau}=\frac{\gamma v}{R}.\label{eq: implies}
\end{equation}

An observer which uses Schwarzschild coordinates measures \cite{key-8}

\begin{equation}
\frac{d\varphi}{dt}=\frac{d\varphi}{d\tau}\frac{d\tau}{dt}=\frac{v}{R}.\label{eq: measures}
\end{equation}

On the other hand, a stationary observer fixed at the radial distance
$R$ measures a proper time related to the Schwarzschild time coordinate
by \cite{key-8,key-12}

\begin{equation}
dt'=\sqrt{1-\frac{2M}{R}}dt.\label{eq: proper time}
\end{equation}

Such a stationary observer also sees a displacement along a circular
arc \cite{key-8}

\begin{equation}
dl=Rd\varphi.\label{eq: arco infinitesimo}
\end{equation}

Then, one can compute the velocity of the particle in orbital motion
as measured by the stationary observer at $R$ like \cite{key-8}

\begin{equation}
\frac{dl}{dt'}=\frac{R}{\sqrt{1-\frac{2M}{R}}}\frac{d\varphi}{dt}=\frac{v}{\sqrt{1-\frac{2M}{R}}}.\label{eq: orbital motion}
\end{equation}

By integrating this expression one gets the time measured by the stationary
observer at $R$ for the particle to travel an arc displacement $\triangle l$
on Earth \cite{key-8}

\begin{equation}
\triangle t'=\frac{\triangle l}{v}\sqrt{1-\frac{2M}{R}}.\label{eq: tempo di viaggio}
\end{equation}

The neutrinos' trajectory can be approximately thought as a circular
arc $\triangle l$, which as a numerical value is 731 kilometers \cite{key-8}.
The author of \cite{key-8} claims that eq. (\ref{eq: tempo di viaggio})
gives the correct answer of the measured time by a stationary observer
when these particles travel an angle distance $\triangle l$ because
the time given by formula (\ref{eq: tempo di viaggio}) is shorter
than the time $\frac{\triangle l}{v}$ the particles will take if
they were traveling in flat space-time. Even if the above analysis
is correct we easily show that the final conclusion is not correct.
In fact, the value of $\frac{2M}{R}$ is of order $6.9*10^{-10}$
\cite{key-4} which gives a correction on $\frac{\triangle l}{v}$
of order $3.4*10^{-10}$. The effect originally measured by the OPERA
Collaboration was of order $10^{-5}$ \cite{key-1}. Therefore the
gravitational effect is five orders of magnitude less than the one
originally measured by OPERA. 

Now, let us make the computation by using the approximate solution
(\ref{eq: sol. approssimata}). Again, we simplify the problem by
assuming a circular orbit, i.e. $r=R,$ $\theta=\frac{\pi}{2}$. If
one sets such values in the coordinates (\ref{eq: sol. approssimata})
the particle travels at the equator, i.e. $R=a_{1}$ \cite{key-9}.
The world line for a massive particle which travels in a circular
orbit $r=R$ at a velocity $v'$ in the spacetime of eq. (\ref{eq: sol. approssimata})
is 

\begin{equation}
x^{\mu}(\tau)=\left(\gamma'\tau,\mbox{ }R,\mbox{ }\frac{\pi}{2},\mbox{ }\frac{\gamma'v'\tau}{R}\right),\label{eq: worldline-1}
\end{equation}

where the condition (\ref{eq: 4-velocity}) now sets the general gamma
factor $\gamma'$ as 
\begin{equation}
\gamma'=\frac{1}{\sqrt{1+\frac{MJ_{2}}{R}+\omega^{2}R^{2}-\frac{v'^{2}}{R^{2}}\left(1+\frac{2M}{R}\right)}}.\label{eq: gamma factor-1}
\end{equation}

Eq. (\ref{eq: worldline-1}) gives 

\begin{equation}
\frac{d\varphi}{d\tau}=\frac{\gamma'v'}{R}.\label{eq: implies-1}
\end{equation}

An observer which uses the coordinates (\ref{eq: sol. approssimata})
measures 

\begin{equation}
\frac{d\varphi}{dt}=\frac{d\varphi}{d\tau}\frac{d\tau}{dt}=\frac{v'}{R}.\label{eq: measures-1}
\end{equation}

Now, a stationary observer fixed at the radial distance $R$ measures
a proper time related to the time coordinate by \cite{key-8,key-12}

\begin{equation}
dt'=\sqrt{1+\frac{MJ_{2}}{R}+\omega^{2}R^{2}}dt.\label{eq: proper time-1}
\end{equation}

Such a stationary observer also sees a displacement along a circular
arc 

\begin{equation}
dl=\sqrt{1+\frac{2M}{R}}Rd\varphi.\label{eq: arco infinitesimo-1}
\end{equation}

Hence, we can compute the velocity of the particle in orbital motion
as measured by the stationary observer at $R$ like 

\begin{equation}
\frac{dl}{dt'}=\frac{\sqrt{1+\frac{2M}{R}}R}{\sqrt{1+\frac{MJ_{2}}{R}+\omega^{2}R^{2}}}\frac{d\varphi}{dt}=\frac{\sqrt{1+\frac{2M}{R}}v'}{\sqrt{1+\frac{MJ_{2}}{R}+\omega^{2}R^{2}}}.\label{eq: orbital motion-1}
\end{equation}

By integrating this expression one gets the time measured by the stationary
observer at $R$ for the particle to travel an arc displacement $\triangle l$
on Earth in the coordinates (\ref{eq: sol. approssimata})

\begin{equation}
\triangle t'=\frac{\triangle l}{v'}\frac{\sqrt{1+\frac{MJ_{2}}{R}+\omega^{2}R^{2}}}{\sqrt{1+\frac{2M}{R}}}.\label{eq: tempo di viaggio-1}
\end{equation}

$\frac{MJ_{2}}{R}$ is order $10^{-13}$ while $\omega^{2}R^{2}$
is order $10^{-12}$ \cite{key-9}. These values, together with the
value of $\frac{2M}{R}$, give a correction on $\triangle l$ again
of order $10^{-10}$ which cannot explain the original OPERA's results
as the effect originally measured by the OPERA Collaboration was of
order $10^{-5}$ \cite{key-1}. 

In summary, in this letter we have shown that the claims of the author
of \cite{key-8} to have very easily solved and explained within the
general theory of relativity that OPERA's neutrinos are not traveling
faster than the speed of light and the early time arrival is due to
the presence of the Earth's gravitational field are not correct. In
fact, the presence of the gravitational field generates a variation
of the proper time of order $10^{-10}$ in both of the Schwarzschild
solution of Einstein\textquoteright{}s field equations and the approximate
solution of Einstein\textquoteright{}s field equations where GPS receivers
are usually synchronized, while the effect originally  measured by
the OPERA Collaboration is of order $10^{-5}$ \cite{key-1}. Thus,
although it looks that data suggesting that neutrinos can travel faster
than light probably resulted from a faulty connection in a GPS timing
system \cite{key-13}, it is important to clarify that, in any case,
the general relativistic effect discussed by Garcia-Islas in \cite{key-8},
cannot explain the original OPERA's data \cite{key-1}.

\paragraph*{Acknowledgements}

The R. M. Santilli Foundation has to be thanked for partially supporting
this letter\emph{ }(Research Grant Number RMS-TH-5735A2310). It is
a pleasure to thank Ruggero M. Santilli, Erasmo Recami, Ammar Sakaji,
Ignazio Licata, Herman Mosquera Cuesta, Gaetano Lambiase and Lawrence
Crowell for various interesting comments and discussions on the issue
of the neutrino faster than the speed of light.

\end{document}